\documentclass[preprint2]{aastex}

\newcommand{\kms}{km~s$^{-1}$}

\begin{document}

\title{A Comprehensive Radio and Optical Study of Abell 2256: Activity from an Infalling Group}
\author{Neal A. Miller\altaffilmark{1}} 
\affil{NASA Goddard Space Flight Center \\ UV/Optical Branch, Code 681\\Greenbelt, MD \ 20771}
\email{nmiller@stis.gsfc.nasa.gov}

\author{Frazer N. Owen}
\affil{National Radio Astronomy Observatory\altaffilmark{2}, P.O. Box O, \\ Socorro, New Mexico 87801}
\email{fowen@aoc.nrao.edu}

\and

\author{John M. Hill}
\affil{Large Binocular Telescope Observatory, University of Arizona \\ Tucson, AZ \ 85721}
\email{jhill@as.arizona.edu}

\altaffiltext{1}{Visiting Astronomer, Kitt Peak National Observatory, National Optical Astronomy Observatories, which is operated by the Association of Universities for Research in Astronomy, Inc. (AURA) under cooperative agreement with the National Science Foundation.}

\altaffiltext{2}{The National Radio Astronomy Observatory is a facility of the National Science Foundation operated under cooperative agreement by Associated Universities, Inc.}

\begin{abstract} 

Abell 2256 is a nearby ($z\approx0.06$), rich cluster of galaxies with fascinating observed properties across a range of wavelengths. Long believed to represent a cluster merger, recent X-ray and optical results have suggested that in addition to the primary cluster and subcluster there is evidence for a third, poorer system. We present wide-field, high sensitivity 1.4 GHz VLA radio observations of Abell 2256 in conjunction with optical imaging and additional spectroscopy. Over 40 cluster radio galaxies are identified, with optical spectroscopy indicating the emission source (star formation or AGN) for most of them. While the overall fraction of galaxies exhibiting radio emission is consistent with a large sample of other nearby clusters, we find an increase in the activity level of galaxies belonging to the third system (hereafter, the ``Group''). Specifically, the Group has relatively more star formation than both the primary cluster and main subcluster. The position of the Group is also coincident with the observed cluster radio relic. We suggest that the Group recently ($\sim$0.3 Gyr) merged with the primary cluster and that this merger, not the ongoing merger of the primary and the main subcluster, might be responsible for many of the unusual radio properties of Abell 2256. Furthermore, the greater star formation activity of the Group suggests that the infall of groups is an important driver of galaxy evolution in clusters.

\end{abstract}

\section{Introduction}

Abell 2256 is an historically well-studied cluster of galaxies. It is a rich cluster \citep[richness class 2,][]{abel1958}, and this fact combined with its proximity and similarity to Coma have resulted in numerous observational studies across many wavelengths. Investigations of the incidence of radio emission in the direction of Abell clusters \citep[including the first such study done by][]{mill1960} included Abell 2256 among detections, and subsequent interferometric observations were able to associate radio emission with individual cluster galaxies in Abell 2256 \citep[e.g.,][to be referred to as BF76]{rudn1976,brid1976}. Several of the radio galaxies were shown to be of an unusual class of objects known as ``head-tail'' sources, in which the ``head'' of the radio emission is coincident with the galaxy while the ``tail'' extends out well past the optical boundary of the galaxy. In addition to such galaxies, the cluster also has diffuse radio emission with a very steep spectrum ($\alpha \sim -1.6$, where $S_\nu \propto \nu^\alpha$). At X-ray wavelengths, \citet{gurs1972} noted the detection of Abell 2256 in the catalogs from the {\it Uhuru} mission, placing it among the brighter X-ray clusters of galaxies. It was consequently a popular target for subsequent X-ray missions, including {\it Einstein, ROSAT, ASCA, BeppoSAX}, and most recently {\it Chandra}. Early work in the optical revealed it to have an unusually large velocity dispersion \citep[1351 \kms,][]{fabe1977}.

The combination of these studies has produced an evolving picture of the nature of Abell 2256. The initial impression of Abell 2256 as a prime example of a very rich, dynamically-relaxed system has been dismissed. An increase in the number of measured cluster galaxy velocities indicated probable substructure \citep{fabr1989}, and {\it ROSAT} resolved the X-ray emission sufficiently to identify two maxima with different temperatures \citep{brie1991}. A merger scenario would also support the radio observations, as diffuse radio emission apparently unassociated with specific galaxies (``halos'' and ``relics'') appear to be related to mergers \citep[e.g.,][]{trib1993,giov2000}. The same is likely true for head-tail radio sources \citep{jone1979}.

Recently, the observational evidence that Abell 2256 represents multiple clusters in the process of merging has been bolstered further. X-ray observations from {\it Chandra} provided much more detail of the infalling subcluster, including its lower temperature than the primary cluster \citep[][hereafter SMMV02]{sun2002}. In fact, these authors noted that the boundary between the infalling subcluster and the primary cluster might be a ``cold front'' \citep{mark2000}, which occurs when the core of a cooler subcluster has not yet been disrupted by the merger. The authors also noted an X-ray ``shoulder'' coincident with a very steep spectrum radio source to the East of the primary cluster's center. This feature may also be associated with some prior merger activity. \citet[][hereafter BLC02]{berr2002} increased the number of measured cluster velocities from 89 to 277. The cluster was best decomposed into three separate components, based on both velocity and positional information. The two larger components apparently correspond to the primary cluster and the infalling subcluster, while a third presumably represents a previously unidentified infalling group (consisting of $\sim10\%$ of the total number of galaxies in the cluster). In summary, Abell 2256 has been shown to be a dynamically active system involving the merger of multiple systems.

Consequently, Abell 2256 presents an excellent opportunity for study. In this paper, we seek to address two general questions. First, can the dynamics of Abell 2256 surmised from the galaxy velocities be related to the radio features of the cluster? While there are indications that features such as radio halos and relics are associated with cluster mergers, the wealth of data for Abell 2256 can potentially elucidate the connection. Second, what is the effect of such dynamical activity on member galaxies? Radio and optical studies of several dynamically active clusters have shown that some possess an excess of star-forming galaxies and AGN over their relaxed cluster counterparts \citep[][referred to subsequently as MO03]{owen1999,morr2002,moA2255}. Is there any evidence for more activity -- either star formation or AGN -- in Abell 2256? What can Abell 2256 tell us about the possible drivers of galaxy evolution in clusters?

In this paper, we present a comprehensive radio and optical investigation of Abell 2256. We have performed wide-field, high sensitivity radio observations of Abell 2256 with the National Radio Astronomy Observatory's (NRAO) Very Large Array (VLA). These observations encompassed the entire cluster, from its core out past the Abell radius (an area of over 2 square degrees), and identified galaxies whose radio emission amounted to $\lesssim2 \times 10^{21}$ W Hz$^{-1}$ and greater. This corresponds to all AGN and galaxies forming stars at rates in excess of 1.2 M$_\sun$ yr$^{-1}$ \citep[using the relation found in][]{yun2001}. The radio data are complemented by new wide-field optical images collected in $R_c$, and optical spectroscopy of many of the radio-selected galaxies. These spectra further increase the velocity database for the cluster, and are used to classify the active galaxies.

The paper is organized as follows. In Section \ref{sec:obs} we give an overview of the observations and data reductions for the radio and optical imaging and the optical spectroscopy. The basic results and analysis are outlined in Section \ref{sec:results}, followed by their implications in Section \ref{sec:discuss}. Conclusions are summarized in Section \ref{sec:conclude}. We have adopted $H_o = 75$ km s$^{-1}$ Mpc$^{-1}$ and $q_o=0.1$, meaning 1\arcsec{} corresponds to 0.98 kpc at Abell 2256.

\section{Observations and Reductions}\label{sec:obs}

\subsection{Radio}\label{sec:radio_obs}

Radio observations at 1.4 GHz were made with the VLA in two configurations. The higher resolution observations were performed during July 1998 while the VLA was in its B configuration, and lower resolution observations were performed during May 1999 in the D configuration. The B configuration provides a resolution of $\sim$5\arcsec{} but is insensitive to structures larger than $\sim$120\arcsec. Abell 2256 includes several extended radio galaxies as well as diffuse emission associated with halo and relic sources \citep[e.g.,][]{rott1994,kim1999}, thereby necessitating the inclusion of the lower resolution D array data.

The observational strategy paralleled that of other wide-field radio surveys \citep[e.g., the NVSS and FIRST;][]{cond1998,beck1995}. A grid of 43 pointings was used to cover the cluster from its core out to large radii ($\gtrsim$50\arcmin, or about 3 Mpc). These pointings were arranged in a hexagonal pattern, with the primary 37 pointings having a grid spacing of 16\arcmin. The remaining 6 pointings formed a smaller hexagon surrounding the central pointing of the main grid. This configuration was adopted to provide near-uniform sensitivity across the cluster, with extra detail in the cluster core. For an excellent overview of the technique, from observation through to construction of the final wide-field image, the reader is directed to Condon et al. All of the observations were made in ``line mode'' using seven channels of 3.125MHz bandwidth for each intermediate frequency and polarization. The integration time for each pointing was 17 minutes while in the B configuration, and 8.5 minutes in D.

Data reduction was performed using NRAO's Astronomical Image Processing System (AIPS), and is described in more detail in MO03. In practice, the most difficult aspect was determining the optimal set of parameters affecting the raw uv data for each pointing. The goal is a consistent beam shape across all pointings, while adding the least noise to the data and maintaining a beam with good imaging properties (i.e., minimal sidelobes). The high declination of Abell 2256 implies an elongated beam whose position angle changes over the course of the observations. Through careful tapering and gridding of the uv data we maintained a beam of $\sim6.9\arcsec \times 5.6\arcsec$ across all pointings, and used a circular restoring beam of equal area (6.3\arcsec) to produce the final maps for each pointing. The final mosaic had an rms noise of about 40 $\mu$Jy beam$^{-1}$ within 1 Mpc of the cluster center, rising to about 60 $\mu$Jy beam$^{-1}$ at the very edges where fewer pointings contribute to the image. These correspond to 5$\sigma$ detections of sources down to radio luminosities of $1.3 \times 10^{21}$ W Hz$^{-1}$ and $1.9 \times 10^{21}$ W Hz$^{-1}$ at Abell 2256. In addition, the multi-resolution clean option in the AIPS task IMAGR was also used to produce an image including the extended radio emission in the cluster. The result for the central 20\arcmin{} of the cluster is shown in Figure \ref{fig-bigmap}.

\placefigure{fig-bigmap}

The AIPS task SAD (``search and destroy'') was used to identify sources in the final mosaic. This task was directed to locate all sources with peak fluxes greater than four times the rms noise level determined over the surrounding $\sim30$\arcmin. SAD fits these sources with Gaussians and notes their positions and fluxes. It also creates a residual map, which was inspected to locate any extended sources not well fit by Gaussians. These were manually added to the lists, with fluxes determined directly by boxing the sources. Each radio source was then inspected, and only those with either peak or integral fluxes greater than five times the local noise (over the surrounding 5\arcmin) were included in the final radio source lists. In total, the radio source list contained about 1300 entries.

\subsection{Optical Imaging}\label{sec:opt_image}

Optical images ($R_c$) of Abell 2256 were obtained during July 1999 using the KPNO 0.9 meter telescope. These were taken with the Mosaic camera, providing a 59\arcmin{} field at a pixel scale of 0.43\arcsec. To cover the spacings between the eight individual chips in Mosaic, we adopted the standard dither sequence of five images per telescope pointing. The wide field of Mosaic allowed Abell 2256 to be imaged well past its Abell radius in only four pointings. A fifth, deeper pointing was centered on the cluster core. The net exposure time was 750 seconds for each of the four shallower images and 1500 seconds for the deeper central pointing. Observations of three Landolt Selected Areas \citep{land1992} were performed at a range of airmass (1.2 to 2.3) to provide photometric calibration.

Data reduction followed the usual procedures, and details may be found in MO03. The astrometric registration of the images is good to about 0.4\arcsec, as determined from the USNO A2.0 catalog \citep{mone2000}. Comparison of the derived magnitudes for the $\sim70$ observed standard stars with their published values suggests the photometry is accurate to within 0.05 magnitudes. Our galaxy magnitudes were derived using the Gunn-Oke metric aperture \citep{gunn1975}, which corresponds to a radius of 13.1 kpc using our adopted cosmology. Optical galaxies brighter than $M^*_R=-20$ \citep[i.e., $m^*_R+2$ using $M^*_R=-22$ from ][]{owen1989} were included in the formal list of candidate cluster radio galaxies (candidate in the sense that they are either cluster members or merely radio galaxies seen in projection), as in \citet{mill2001}. This corresponds to $m_{R_c}=16.92$ for Abell 2256, adjusted for $A_{R_c}=0.14$. The SExtractor software package \citep{bert1996} was used to identify all objects in the fields as well as classify them and calculate their Gunn-Oke magnitudes. Finally, the images were visually inspected to confirm the star/galaxy segregations made by SExtractor.

The list of optical galaxies was then correlated with the list of radio sources. For this procedure, we adopt a maximum probability for chance coincidence of radio and optical sources of $0.5\%$ \citep[for a brief description, see][]{mill2001}. For Abell 2256, this amounted to a maximum positional separation of just over 10\arcsec. Three galaxies with separations $>$8\arcsec{} were removed from the list upon inspection when their radio sources were clearly associated with fainter background galaxies. This meant that all candidate cluster radio galaxies had radio-optical separations of 5\arcsec{} or less. Two extended radio galaxies (the tailed radio galaxies J170330+783755 and J170330+783955) were manually added to the list (see Figures \ref{fig-koalatad} and \ref{fig-flagpole}). The resulting list contained 54 candidate cluster radio galaxies brighter than $M_{R_c}=-20$ and within 3 Mpc of 17:04:02 +78:37:55 (J2000), our adopted center coordinate of Abell 2256.

\placefigure{fig-koalatad}
\placefigure{fig-flagpole}

\subsection{Optical Spectroscopy}

Optical spectra of galaxies in Abell 2256 were obtained during two separate observing runs in 1999. In May 1999, 18 candidate cluster radio galaxies were observed using the Kitt Peak 2.1-meter telescope and the GoldCam spectrograph. Using a 2\arcsec{} slit and a 300 line mm$^{-1}$ grating, spectra covering 3700--8000$\mbox{\AA}$ at $\sim 7\mbox{\AA}$ resolution were obtained. The integration time was 900 seconds, and wavelength calibration was achieved using He-Ne-Ar lamp exposures taken after the science exposures. Details may be found in \citet{mill2001}. Numerous cluster galaxies (including non-radio galaxies) were also observed using the MX Spectrometer \citep{hill1986,hill1988} and the Steward Observatory Bok Telescope (2.3m). This system uses 2\arcsec{} fibers on movable probes to collect up to 32 galaxy spectra in a 45\arcmin{} field, along with 30 sky spectra from fibers ``piggy-backed'' to the target probes. A 400 line mm$^{-1}$ grating provided wavelength coverage from 4000$\mbox{\AA}$ to 7300$\mbox{\AA}$ at 7$\mbox{\AA}$ resolution. Five pointings each of one-hour duration were made. The cluster Abell 2255 was also observed during this run, and details for the observations and determination of velocities may be found in \citet{hill2003} which addresses the kinematics of that cluster.

\placetable{tbl-galvel}

We report velocity measurements for 73 galaxies in Table \ref{tbl-galvel}. Many of these have been published previously in the work done by \citet{fabr1989} and BLC02, but for 22 galaxies the velocities are new. Most of these new velocity measurements are at the periphery of the cluster, and in particular to the Northwest along the extension of the line connecting the two X-ray peaks. In addition to velocity measurement, we classified the radio galaxies as star forming or AGN on the basis of their emission and absorption features. This was achieved using the standard emission line ratio diagnostics \citep[e.g.,][]{bald1981,veil1987} after correcting for stellar Balmer absorption. Our procedures are described in detail in \citet{mill2002}. Unfortunately, the wavelength coverage of the MX spectra did not include [OII] for cluster galaxies, preventing us from applying the MORPHS classification scheme to many of the cluster radio galaxies \citep[see][for a description of this classification scheme]{dres1999}.

Our velocities and those of BLC02 are generally consistent, with our velocities being $-42$ \kms{} lower on average. As this is within the observational errors, we have not applied any zero point offsets to the velocities in producing a combined velocity database. The cluster includes 294 cluster members, as determined through 3$\sigma$ clipping. The systemic velocity was found to be $17490 \pm 74$ \kms, with a velocity dispersion of $1269^{+56}_{-49}$ \kms{} \citep[using the biweight estimators of][]{beer1990}. These values were adopted to determine whether the identified radio galaxies were cluster members; that is, we have accepted all galaxies in the range $13683 - 21297$ \kms{} as cluster members.

The completeness of our velocity database was determined as a function of both optical magnitude and radial separation from the cluster center, and is depicted in Figure \ref{fig-vcomp}. For the entire cluster out to 3 Mpc in radial extent, we have measured velocities for $70\%$ of those galaxies brighter than the formal $M^*+2$ limit. A large fraction of the galaxies without velocity measurements are fainter galaxies at the larger radial distances of the study. Ulabeledsing a 2 Mpc radial limit and ignoring galaxies fainter than $M^* + 1.5$, the velocity completeness is about $90\%$. Note that there are an additional 74 galaxies fainter than $M^*+2$ for which velocities were obtained.

\placefigure{fig-vcomp}

\section{Results}\label{sec:results}
\subsection{Overview of the Radio Source Population}

Using our data supplemented with that of BLC02, 49 of the 54 candidate cluster radio galaxies have measured velocities. Of these, 40 are confirmed cluster members. The list of these galaxies and their attributes (velocities, magnitudes, radio fluxes, and classifications) may be found in Table \ref{tbl-clusRG}. Prior papers \citep[BF76;][]{brid1979,rott1994} applied alphabetical designations to the prominent radio features, and we have included these designations in Table \ref{tbl-clusRG} when they are associated with specific galaxies. In addition, the alphabetical designations have been noted in Figure \ref{fig-bigmap}. To avoid confusion we will adhere to the following terms and definitions. The radio {\it halo} in Abell 2256 is the diffuse emission around the cluster center. Its surface brightness is too low for it to be present in our images, although it may be associated with one cluster radio galaxy ( ``D'' in the prior studies). The {\it relic} is the diffuse emission to the North of the cluster center, identified as ``G'' and ``H.'' It is often referred to as a halo, although under current usage of the term this is incorrect as it has a high polarization \citep[up to $50\%$;][]{clar2001} and is located off the cluster center. Lastly, the diffuse emission labeled ``F'' will be referred to as the ultrasteep spectrum source \citep{brid1979,rott1994}. It is most likely another relic source.

\placetable{tbl-clusRG}

Our spectra include 28 of the 40 confirmed cluster radio galaxies, an even split of which are powered by each AGN and star formation. Among the AGN, we associate four galaxies with head-tail radio sources. \citet{brid1979} identified these four galaxies as definite head-tails (features A, B, C, and I; see Table \ref{tbl-clusRG} and Figures \ref{fig-koalatad}, \ref{fig-flagpole}, and \ref{fig-flag}), and noted four additional potential head-tails (F, G, H, and M). The resolution of our maps excludes source M as a cluster radio galaxy, and source F (the ultrasteep spectrum source; see Figure \ref{fig-f}) is unlikely to be a true head-tail source. It is resolved into three components by our observations (potentially attached with lower surface brightness emission than our observations detect), the easternmost of which (F3) is associated with a cluster galaxy \citep[and probably unassociated with the remainder of the emission of source F,][]{rott1994}. Sources G and H comprise the relic source, with Source G apparently associated with a cluster elliptical, as the peak in the radio emission fell only 3.1\arcsec{} from the galaxy (see Figure \ref{fig-relic1}). This galaxy may have once been a head-tail radio galaxy which was disrupted by the merger environment \citep[as suggested by][and further explored in Section \ref{sec:discuss}]{rott1994}. Even if only four of the galaxies are associated with head-tail radio sources, this is an unusually high number for a cluster. Typical clusters only possess from 0--2 such sources \citep{ledl1995}.

One further cluster AGN may be associated with extended emission. The galaxy located at J170448+783829 is one of the brighter ellipticals in Abell 2256 and is associated with an unresolved 11.4 mJy radio source (radio feature D in BF76; see Figure \ref{fig-center}). It sits in the center of the diffuse cluster halo emission \citep{rott1994}, and about 1\arcmin{} to the East of the brightest cluster galaxy (which is not a radio source). Of the remaining cluster AGN for which we have optical spectra, all but one are low-luminosity radio sources with absorption line spectra. The spectrum of J170150+790216 is that of a strong LINER. The radio emission is unresolved, and the optical image of the galaxy suggests a recent merger or interaction (see Figure \ref{fig-fig8}).

\placefigure{fig-flag}
\placefigure{fig-f}
\placefigure{fig-relic1}
\placefigure{fig-center}
\placefigure{fig-fig8}

Fourteen cluster radio galaxies are confirmed to be star-forming on the basis of their spectra. The wavelength coverage adopted for the multi-fiber observations did not include cluster [OII] emission, preventing us from classifying the galaxies in the MORPHS scheme. However, we can compensate for this somewhat by assuming that the equivalent width of [OII] is $40\%$ that of the H$\alpha$-[NII] equivalent width \citep{kenn1992}. Making this assumption, we find that one of the galaxies (J170534+785116) is a starburst galaxy, or ``e(b)'' in the MORPHS classification scheme. Not coincidentally, this galaxy has the largest 1.4 GHz flux of all the cluster star-forming galaxies ($\sim$4.2 mJy) and was even identified in the early radio study of BF76 (source ``E''). Three of the star-forming galaxies have H$\delta$ absorption sufficient enough to place them in the ``e(a)'' class (J165253+784629, J170335+785055, and J171016+780207). Galaxies with such spectra are likely associated with starbursts in which selective extinction has diminished their [OII] emission in dusty HII regions, yet longer-lived A stars have exited (or dispersed) these HII regions and produce a strong H$\delta$ absorption feature \citep{pogg1999}. One further galaxy has H$\delta$ absorption which places it close to the e(a) category (J170512+784814). The remainder of the star-forming galaxies appear to be relatively normal, given the limits of our data.

Six of the remaining galaxies with measured velocities are background radio galaxies seen in projection on the cluster, and the remaining three are foreground radio galaxies. These galaxies, along with the five candidate cluster radio galaxies which lack velocities, are listed in Table \ref{tbl-moreRG}. Relaxing the formal limits of $M_R \leq -20$ and the radial limit of 3 Mpc produces a further four radio galaxies with measured velocities which are also included in Table \ref{tbl-moreRG}. Three of these are cluster members, while the fourth is a background galaxy.

\placetable{tbl-moreRG}

BLC02 note that most of the non-cluster galaxies in their velocity list come from a background group at a velocity of 52,413 \kms. Three of our background radio galaxies appear to belong to this system, including an apparent FR I source (J170454+782954, see Figure \ref{fig-backfr1}). This source has a radio luminosity of $10^{23}$ W Hz$^{-1}$ and optical magnitude of $-22.5$, typical of FR I sources with wide-angle tail morphology. Another obvious radio galaxy can be seen in the figure (J170515+783018), although the velocity of this galaxy is unknown. Should it belong to the same system, it would have a radio luminosity of $\sim4 \times 10^{23}$ W Hz$^{-1}$ and an absolute magnitude of $-21.6$. Lastly, there is faint, diffuse emission just to the North of the FR I source. Some of this may be associated with the faint galaxy at the western edge of the emission (J170450+783011), although it is unlikely that this represents a head-tail galaxy at the redshift of the FR I source. If the galaxy were in the same system, its absolute magnitude of $-20.7$ would be unusually faint for a head-tail source.

\placefigure{fig-backfr1}

\subsection{Comparison to Other Clusters}

How does the radio galaxy population of Abell 2256 compare to other nearby clusters? In MO03 this question was investigated for 20 nearby clusters, including Abell 2256. The analysis used a chi-square test to evaluate the fraction of galaxies with radio emission in a given cluster compared to the same fraction for the composite sample formed by the other 19 clusters. In that paper, the dynamically active cluster Abell 2255 was demonstrated to have a highly significant excess of radio galaxies, but Abell 2256 appeared relatively normal. This result is not affected by the lack of velocities for five potential members of Abell 2256; three of these have radio luminosities lower than the limit adopted for the other clusters, and the remaining two do not change any of the statistical results. We note that the lack of an excess of radio galaxies in Abell 2256 is consistent with the low blue fraction ($f_B = 0.03 \pm 0.01$) for the cluster, reported in \citet{butc1984}.

There is one specific case in which the radio galaxy population of Abell 2256 differs from the other clusters, and that is the fraction of optically brightest galaxies ($M_R\leq-22$) harboring radio sources. There are seven such radio galaxies in Abell 2256, six of which have spectra indicative of AGN (five have absorption line spectra, the sixth is a strong LINER). While we do not have a spectrum for the seventh galaxy, it is probably also an AGN. It is the middle galaxy in a triple galaxy system, the western galaxy of which is a very strong tailed radio galaxy (see Figure \ref{fig-koalatad}). The excess of these optically-bright radio galaxies is not unexpected, as many prior studies have noted the unusually large number of head-tail radio galaxies in Abell 2256. In addition, the {\it Chandra} observations of SMMV02 noted at least three X-ray point sources corresponding to cluster elliptical galaxies.

\subsection{Radio Galaxy Positions and Relation to Substructures}

Abell 2256 is known to contain substructure. BLC02 used the KMM algorithm \citep[``Kaye's mixture model,'' see][]{ashm1994} to decompose the cluster into three separate substructures based on both velocity and position. This is a maximum likelihood algorithm, which fits a specified number of Gaussians to the data and determines to which of these Gaussians each individual galaxy most likely belongs. Roughly 70$\%$ of the galaxies belonged to the primary cluster (Gaussian \# 2 in BLC02, referred to hereafter as the ``Primary''), and about 20$\%$ belonged a second subcluster (Gaussian \# 1, hereafter the ``Subcluster''), presumably the infalling subcluster identified in the X-ray observations. The remaining $\sim10\%$ belonged to a third group (Gaussian \# 3, hereafter the ``Group'') concentrated to the North of the Primary center and at higher radial velocity. 

We have repeated the KMM procedure on our combined velocity list in order to place all galaxies (including the radio galaxies) into their respective substructures. Our KMM code differs slightly from that of \citet{ashm1994}, in that we assume the covariances among the three attributes (two position, one velocity) are zero.\footnote{The presence of real substructure actually implies some correlation between position and velocity, or non-zero covariance. In the case of Abell 2256, application of the KMM algorithm used by \citet{ashm1994} resulted in identical assignments for nearly $98\%$ of all cluster galaxies.} Thus, we calculate the distance from each galaxy to each specified Gaussian, $i$:
\begin{equation}
f_i = \exp \left[ -\left( \frac{\alpha - \bar\alpha_i}{2 \sigma_{\alpha_i}}\right) ^2 - \left( \frac{\delta - \bar\delta_i}{2 \sigma_{\delta_i}}\right) ^2 - \left( \frac{v - \bar v_i}{2 \sigma_{v_i}}\right) ^2 \right].
\end{equation}
Each galaxy is assigned to the Gaussian which produces the smallest distance, and once all galaxies have been assigned the new means and dispersions are calculated. These values are then used in successive iterations until the assignments are stable. In particular, the log likelihood of the fit is described by:
\begin{equation}
L_C = \sum_{i=1}^{ng} \sum_{j=1}^{N} z_{ij}[\ln \pi_i + \ln f_i ]
\end{equation}
where the summation over $i$ refers to the number of groups to be fit to the data and the summation over $j$ refers to the individual galaxies. The factor $z_{ij}$ refers to the probability that a given galaxy $j$ belongs to group $i$, which is evaluated based on the values of $f_i$ for that galaxy. The factor $\pi_i$ is the ``mixing fraction,'' or the fraction of the total population which are members of group $i$. Convergence of the algorithm is assumed once $L_C$ has converged. As discussed in \citet{ashm1994}, the significance of a multi-component fit is not easily surmised. Consequently, our code also calculates $L_C$ for Monte Carlo shuffles of the data in order to determine the frequency with which the algorithm can fit a randomized data set as well as it does the real data. In this case, we only shuffled the velocity data so as to avoid issues related to positional sampling of cluster velocities (our MC test is thus directly analogous to that used to quantify the significance of the Dressler-Shectman test). The results are presented in Table \ref{tbl-kmm}; the derived velocity means and dispersions are consistent with those presented in Table 8 of BLC02. In total, 186 galaxies were assigned to the Primary, 78 galaxies to the large infalling Subcluster, and 30 galaxies to the Group identified primarily on the basis of its higher radial velocity. Only $2.5\%$ of the MC shuffles had 3-Gaussian fits as good as or better than the actual data. Note that for the two-component and four-component fits the results were $17.0\%$ and $4.1\%$, respectively, arguing for the three-component model (when attempting a fifth component, the algorithm failed to converge without removing nearly all of the galaxies from one of the substructures). 

\placetable{tbl-kmm}

A relatively large number of the radio sources were assigned to the Group (7 of 30 galaxies with velocities placing them in this structure are in Table \ref{tbl-clusRG}). Spatially, this may be seen in Figure \ref{fig-dist} as an apparent excess of radio galaxies to the North of the cluster center. Quantitatively, a chi-square test places this excess of radio sources at only marginal significance, about $90\%$. Individually, there is no statistical difference in the radio galaxy fractions of the Primary and the Subcluster. If we include the three additional confirmed cluster radio galaxies which were fainter than our $M_R=-20$ limit (Table \ref{tbl-moreRG}), the excess of radio sources in the Group is accentuated. One of these sources belongs to the Group, increasing the significance of the radio galaxy excess to over $95\%$.

\placefigure{fig-dist}

Interestingly, the radio galaxies in the Group are almost exclusively associated with star formation. Our spectra include six of the eight identified radio galaxies in the Group, all of which exhibit emission lines. One source, J170150+790216, is a strong LINER whose optical morphology suggests a recent galaxy-galaxy interaction (see Figure \ref{fig-fig8}). This galaxy is also separated from the other radio detections of the Group, lying at a higher declination. The remaining five galaxies have spectra of star-forming galaxies. The propensity for the Group members to be radio-detected star-forming galaxies is significant at nearly $99.9\%$ (5 of 30 compared to 9 of 264), although this claim is tempered by our incomplete spectroscopy. However, should all of the Group galaxies lacking spectroscopic classification be AGN and all the non-Group galaxies lacking spectroscopic classification be star forming (i.e., a worst case scenario), the excess of star-forming galaxies in the Group would still be significant at over $94\%$. Star formation therefore seems more prevalent in the Group than in the remainder of Abell 2256. Additionally, we note that the two Group member radio galaxies for which we do not possess spectra appear to be spirals and their radio-to-optical flux ratios are suggestive of star formation rather than AGN. 

Conversely, only one of the radio galaxies assigned to the Subcluster is associated with star formation. Our KMM code suggests that the probability that this galaxy belongs to the Subcluster is only slightly greater than the probability that it belongs to the Primary. Thus, the activity level of the infalling Subcluster appears to be quite low.

\section{Discussion}\label{sec:discuss}

Abell 2256 is clearly a dynamically complicated system. Attempts to understand its properties have generally relied upon the notion that a major merger was occurring, and have therefore focussed on the interaction of the primary cluster with the large subcluster to the West. However, recent evidence has revealed the presence of a third group. This group was identified in the large number of cluster galaxy velocities obtained by BLC02, and SMMV02 found X-ray evidence for a group which presumably had already merged with the primary cluster. We will argue for the importance of this third substructure in understanding some of the radio properties of Abell 2256. 

The key pieces of evidence for this interpretation are depicted in Figure \ref{fig-relicpos}. Here, the radio relic is shown along with the positions of the Group member galaxies. It is remarkable that seven of the eight radio galaxies belonging to the Group are located immediately to the North of the relic. With the exception of the two galaxies for which we do not have optical spectra (J170322+784655 and J170455+782954), all of these Group radio galaxies are powered by star formation. In addition, three further galaxies were assigned to the Primary by the KMM algorithm but are arguably associated with the Group. The algorithm assigned marginally higher probabilities that J170335+785055, J170448+783829, and J170628+784915 were associated with the Primary than the Group. The first and third of these are star-forming galaxies (one of which is slightly fainter than our formal $M_R=-20$ limit) located among the other Group star-forming galaxies. The second is the fairly strong radio galaxy located at the rough center of the radio halo (see also Figure \ref{fig-center}). It is flanked by several fainter galaxies which were assigned to the Group. Thus, it appears that the galaxy activity associated with the Group may be even larger than previously stated.

\placefigure{fig-relicpos}

Based on this positional evidence, we propose that the Group may be responsible for the radio relic source. This was also suggested by BLC02, who noted the higher velocity of the Group and placed it between the observer and the Primary and currently falling into the cluster. Should the merger of the Group with the Primary be responsible for the radio relic, the Group is more likely {\it behind} the Primary having recently plunged through the core of the Primary. In fact, SMMV02 argued that should the X-ray ``shoulder'' identified in the {\it Chandra} image be associated with a group, models suggest that the system is viewed about 0.3 Gyr after core passage. The passage of the Group through the intracluster gas would create a shock, which in turn has powered the relic source. This can also explain the halo, which is centered more on the high-velocity galaxy J170448+783829 \citep[this possibility has also been noted in][]{clar2001}. It is also interesting to note that the models of \citet{roet1995} suggested that Abell 2256 was better described by two mergers, one of which occurred $\sim0.5$ Gyr previously. Oddly enough, the premise for this result was the recreation of structure indicated in the {\it ROSAT} X-ray temperature maps \citep{brie1994}, an observational finding which was not confirmed by the later satellite observations \citep[{\it ASCA}, {\it BeppoSAX}, and {\it Chandra};][]{mark1996,mole2000,sun2002}. 

Several parameters of the relic source are potentially explained by this geometry and scenario. \citet{brid1979} noted the flat spectral index of the radio emission from the relic, and used this to estimate an age of $\sim 3 \times 10^7$ years. A young age may also be implied by the morphology of the relic -- in the model of \citet{enss2002}, ``sheet-like'' relics are younger and have flatter spectral indices. In this model, relics are created when ``fossil radio cocoons'' are adiabatically compressed by a shock front created during a merger. These cocoons are formed by now-defunct radio galaxies which filled small volumes within the intracluster gas with hotter, more diffuse gas and magnetic fields. The emitting electrons remain in these cocoons, having faded from synchrotron emission. The compression increases the energy of the cocoons and increases their magnetic field strengths, thereby once again increasing their synchrotron emission and producing radio relic sources. If we assume this model for Abell 2256, it is possible that the galaxy associated with the peak of the relic emission (Figure \ref{fig-relic1}) was the progenitor of the cocoon which formed the relic. At early ages, relics produced in this manner can exhibit loop-like structures akin to what is seen along the Eastern edge of the relic in Abell 2256, depending on viewing geometry. Lastly, the model also predicts polarizations like that in the Abell 2256 relic. Given the morphology and polarization of the relic, the Group would be moving largely along our line of sight but inclined slightly such that it entered from the South. Performing a simple calculation wherein this inclination angle is 45$^\circ$, the separation between the star-forming Group galaxies and the cluster center implies a travel time of $\sim$0.3 Gyr at the relative velocity of the Group, consistent with the estimate from the X-ray observations.

One possible difficulty with this model is the relatively low velocity dispersion of the Group. The identified Group has a velocity dispersion of around 300 \kms, which is consistent with typical values for rich groups \citep[e.g.,][]{ledl1996}. N-body simulations have demonstrated that groups are effectively dispersed as they pass through the cores of more massive clusters \citep[e.g.,][]{roet1993,pink1996}, thereby increasing their velocity dispersions. Thus, we would expect a larger velocity dispersion for the Group should it be viewed after rather than before core passage. Obtaining the true velocity dispersion for the Group would require perfect knowledge of the assignments of galaxies to their pre-merger substructures, information which is not available. The low velocity dispersion is apparently a consequence of how the galaxies are assigned. In Figure \ref{fig-trump} we plot the velocity difference between each galaxy and the velocity center of its associated substructure as a function of radial separation from the center of that substructure. The assignments are heavily influenced by velocity; the ``trumpet'' shape for a virialized system is not easily apparent for any of the three fitted systems. Apparently, only the higher velocity members of the initial group are assigned Group membership by the KMM algorithm, thus reducing the fitted dispersion of this substructure. On the other hand, the positions of the Group members do resemble the results of N-body simulations viewed shortly after core passage. The Group members become more dispersed as one moves along the apparent axis of the merger from the center of the cluster to the Northwest (see Figure \ref{fig-relicpos}).

\placefigure{fig-trump}

Another mystery of the radio population in Abell 2256 is the extreme length and straightness of the tailed radio galaxy J170330+783955 (Figure \ref{fig-flagpole}). The tail extends about 300 kpc in our image, although sensitive observations at lower resolution suggest it is potentially twice as long as this \citep{rott1994}. Since the emission is not disrupted, the region must have very low turbulence -- a requirement which is contrary to the generation of a relic source. However, in the discussion above the relic would be located in the back half of the cluster which would presumably require the head-tail radio galaxy to lie in front of it. It is also likely that the Subcluster is entering the Primary slightly from behind (BLC02,SMMV02).

Evidence for the Group being seen shortly post-merger might also be found in the spectra of Group members. \citet{bekk1999} modeled the effect of a cluster-group merger on the star formation history of a disk galaxy initially associated with the infalling group. The time-dependent tidal gravitational field of the merger was able to induce a secondary burst of star formation in the galaxy. The merger also removes a significant fraction of the disk galaxy's halo, thereby reducing its reservoir of gas to replenish the disk. Other cluster environmental effects run parallel to this; ram pressure stripping should also effectively remove gas from infalling galaxies \citep[e.g.,][]{quil2000}. In total, we might expect to see galaxies with starburst and post-starburst features among the spectra of the Group members in Abell 2256. For example, this seems to be the case in the Coma cluster as post-starburst signatures (i.e., strong Balmer absorption) are seen in several galaxies associated with the NGC 4839 system \citep{cald1993,bekk1999}, which is believed to have recently crossed through the core of the Coma cluster \citep{burn1994}. Similarly, in distant clusters a population of starbursts in which selective dust extinction produces strong post-starburst features along with muted star formation features (i.e., [OII] emission) has been observed \citep[the e(a) galaxies of][]{dres1999,pogg1999}. In some cases, the dust extinction may be severe enough to remove all emission lines and produce a post-starburst spectrum despite ongoing star formation \citep{smai1999}. There is slight evidence for such an effect in the Group members. The lone identified starburst galaxy in Abell 2256 belongs to the Group, and one of the three identified e(a) galaxies is the Group outlier J170335+785055. Additionally, the star-forming galaxy with H$\delta$ absorption just less than the formal limit of the e(a) class is a Group member. Unfortunately, the signal-to-noise of our spectra and the frequent lack of [OII] in the spectral coverage makes a rigorous evaluation difficult. For the MX observations, the average SN per pixel evaluated over the regions surrounding H$\delta$ is 6.8, with a minimum of 3.7 and a maximum of 10.9. Improved future spectroscopy would clearly be of great use.

It is also interesting to note the lack of star-forming galaxies in the Subcluster. Only one radio-identified star-forming galaxy was associated with the Subcluster, and our KMM code did not place it in the Subcluster with high confidence. The galaxy lies along the Western edge of the spectroscopically-confirmed cluster members (J165306+783013), and should it belong to the Subcluster it trails the core of this system by 1.7 Mpc in projected distance. It is possible that this galaxy is just an infalling galaxy unassociated with the Subcluster but assigned to it primarily on the basis of its low radial velocity. The dearth of star-forming galaxies does not appear to be limited to those identified through radio selection, either. Our spectroscopy database includes eight cluster emission-line galaxies which were not radio detections. In general, these galaxies lie at the periphery of the cluster where the sensitivity of the radio images is lower, thereby explaining their absence from our radio lists. All of these galaxies were assigned to the Primary by the KMM code. As the velocity dispersion of cluster star-forming galaxies is generally larger than the velocity dispersion of cluster ellipticals and S0's, one might expect that the higher radial velocity star-forming galaxies initially associated with the Subcluster have been assigned to the Primary. However, inspection of the velocity histogram of all galaxies does not indicate that any apparent low radial velocity galaxies were clipped from the overall cluster list. Consequently, it appears that the lack of star-forming galaxies associated with the Subcluster is real.

What is the relationship of this proposed scenario to the findings for other clusters? Of the twenty nearby clusters we have studied, Abell 2255 has the most dramatic increase in galaxy activity. In particular, this cluster has a marked excess of optically-fainter ($M_R > -21$) star-forming galaxies. There are some similarities to its dynamical environment with that of Abell 2256, including evidence for a recent merger between the primary cluster and an infalling cluster/group \citep[with roughly a 8:1 mass ratio;][]{burn1995} and the presence of numerous outlying groups \citep{hill2003}. MO03 argued that a recent merger, with core passage having occurred about 0.1 Gyr previously, was responsible for the increase in the number of cluster star-forming galaxies. The mass ratio between the Primary and the Group in Abell 2256 must be similar to that proposed for Abell 2255, and in each case it appears that core passage apparently occurred relatively recently. The Abell 2256 Subcluster, however, is larger and seemingly has very few (if any) star-forming galaxies which could undergo a burst of star formation. Prior to the merger it was likely a fairly isolated, relaxed cluster. Any star-forming galaxies associated with it would have simply been nearby field galaxies in the process of infall. Presently, such infalling galaxies would more likely be assigned to the Primary. Hence, the merger of the Subcluster with the Primary may have very little effect on cluster-wide star formation as it is not likely to introduce a large number of galaxies which might undergo starbursts. 

Consequently, the acquisition of peripheral groups may be more important in understanding galaxy activity in clusters than mergers of larger partners. Although the simulations of \citet{bekk1999} suggested that larger mass ratios translated to stronger starbursts during cluster mergers, this was based on the assumption that gas-rich spiral galaxies existed in one of the pre-merger partners. As we have noted, the star-forming galaxy population of the rich Subcluster in Abell 2256 is quite small, suggesting little available material for star formation induced by the merger. Simulations of hierarchical growth of structure also indicate that mergers of comparable-mass partners are much rarer than cluster-group mergers, so increased cluster activity as a function of redshift \citep[e.g.][]{butc1984} can more easily be related to the build up of rich cluster through acquisition of outlying groups.

\section{Conclusions}\label{sec:conclude}

We have presented a multiwavelength observational study of Abell 2256. Our deep radio continuum imaging reached an rms sensitivity as low as $\sim35$ $\mu$Jy beam$^{-1}$ and encompassed an area from the center of the cluster out to radii of $\sim$50\arcmin, or about 3 Mpc. Potential cluster radio galaxies were identifed using optical imaging of the same area, with membership being evaluated on the basis of both new and existing optical spectroscopy. In total, 43 cluster radio galaxies were identified and many of these were characterized as AGN or star-forming galaxies on the basis of their spectra. Paralleling the study of BLC02, the application of a maximum likelihood algorithm assigned the galaxies to their probable substructures, which correspond to the Primary, the infalling Subcluster, and a Group.

Taken collectively, the radio galaxy population of Abell 2256 is not unusual when compared to a sample of nineteen other nearby clusters. We confirm a statistical excess of the optically brightest radio galaxies, which might have been guessed from the previously noted unusual number of galaxies with extended radio emission in the cluster. Outside of this specific population, the radio galaxy fraction of Abell 2256 is relatively normal. We do, however, note that the galaxies belonging to the Group are more frequently associated with radio emission and nearly all of these galaxies are powered by star formation. The radial velocity of the Group is $\sim2000$ \kms{} greater than that of the Primary, and its position is coincident with the cluster radio relic source. We therefore argue that the Group has recently merged with the cluster and is located behind the cluster relative to the observer, roughly 0.3 Gyr after core passage. Under this scenario, the radio relic would owe its existence to the cluster-group merger rather than the ongoing cluster-subcluster merger, which appears to be about 0.2 Gyr prior to core passage and entering the Primary from behind and to the West \citep{roet1995}.

In the broader picture, if Abell 2256 is a representative example then galaxy activity in clusters may be most influenced by the merger history of the parent cluster with outlying groups. These minor mergers would provide a relatively large influx of gas-rich spirals which might undergo starbursts followed by truncation of star formation. Mergers of larger partners, while having more profound effects on the ICM, may not prove overly important toward understanding star formation in clusters simply because the partners are less likely to include the potential starbursting galaxies. Thus, active clusters such as those identified in Butcher-Oemler type studies would simply correspond to dynamically young clusters in larger-scale structures rich in peripheral groups.

\acknowledgments

NAM thanks the National Radio Astronomy Observatory for a predoctoral fellowship, and acknowledges the support of a National Research Council Associateship Award at NASA's Goddard Space Flight Center. This research was supported in part by NASA through the American Astronomical Society's Small Research Grant Program. We thank Keith Ashman for providing the original KMM code, and Eric Greisen for assistance with AIPS and the multi-resolution clean.

\clearpage

\begin{deluxetable}{l l r r c r r}
\tablecolumns{7}
\tablecaption{Galaxy Velocity Measurements\label{tbl-galvel}}
\tablewidth{0pt}
\tablenum{1}
\tablehead{
\multicolumn{2}{c}{} & \multicolumn{3}{c}{This Study} &
\multicolumn{2}{c}{Berrington et al.} \\
\colhead{RA} & \colhead{Dec} & \colhead{$cz$} & 
\colhead{$\delta cz$} & \colhead{Comment} & \colhead{$cz$} & 
\colhead{$\delta cz$} \\
\colhead{(J2000)} & \colhead{(J2000)} &
\multicolumn{2}{c}{[km s$^{-1}$]} & \colhead{} & 
\multicolumn{2}{c}{[km s$^{-1}$]}
}
\startdata
16:47:57.9 & 78:52:07 & 17784 & 31 & Emi & \nodata & \nodata \\
16:50:39.6 & 78:39:07 & 16756 & 42 & Abs & \nodata & \nodata \\
16:51:11.5 & 79:06:05 & 17671 & 58 & Abs & \nodata & \nodata \\
16:52:32.0 & 78:57:03 & 30106 & 78 & Emi & \nodata & \nodata \\
16:52:52.8 & 78:46:29 & 16836 & 19 & Emi & \nodata & \nodata \\
16:53:05.9 & 78:30:13 & 16663 & 69 & Emi & \nodata & \nodata \\
16:53:13.6 & 78:44:13 & 16397 & 41 & Abs & \nodata & \nodata \\
16:53:33.3 & 79:00:00 & 18742 & 37 & Emi & \nodata & \nodata \\
16:53:34.3 & 79:06:21 & 18500 & 40 & Emi & \nodata & \nodata \\
16:54:06.1 & 78:37:42 & 16674 & 39 & Abs & 16710 & 36 \\
16:54:16.6 & 79:11:20 & 53488 & 53 & Abs & \nodata & \nodata \\
16:54:29.4 & 78:54:34 & 16228 & 39 & Abs & \nodata & \nodata \\
16:54:36.1 & 79:14:56 & 39573 & 60 & Emi & \nodata & \nodata \\
16:54:58.9 & 79:08:45 & 17433 & 61 & Abs & \nodata & \nodata \\
16:55:32.2 & 78:32:02 & 15969 & 57 & Abs & 16180 & 35 \\
16:55:36.9 & 79:05:10 & 17808 & 30 & Abs & \nodata & \nodata \\
16:56:06.9 & 79:23:28 & 26352 & 60 & Emi & \nodata & \nodata \\
16:56:09.6 & 78:56:36 & 17815 & 46 & Abs & 17851 & 46 \\
16:56:16.1 & 78:55:09 & 18288 & 32 & Emi & 18309 & 48 \\
16:56:18.3 & 79:09:29 & 17598 & 60 & Emi & \nodata & \nodata \\
16:57:05.5 & 78:32:48 & 16350 & 56 & Abs & 16495 & 31 \\
16:57:58.5 & 78:36:15 &  4260 & 22 & Emi &  4253 & 45 \\
16:58:21.1 & 78:43:25 & 15791 & 67 & Abs & 15909 & 52 \\
16:58:46.8 & 78:25:21 & 15632 & 62 & Abs & 15635 & 49 \\
17:00:07.5 & 78:49:52 & 16514 & 68 & Abs & \nodata & \nodata \\
17:00:11.1 & 79:00:16 & 17383 & 70 & Abs & 17415 & 70 \\
17:00:11.5 & 79:03:33 & 17597 & 58 & Abs & 17668 & 34 \\
17:00:16.0 & 78:34:25 & 18362 & 47 & Abs & 18355 & 62 \\
17:00:20.4 & 78:33:55 & 16454 & 62 & Abs & 16372 & 55 \\
17:00:32.6 & 78:56:10 & 15021 & 65 & Abs & 15237 & 47 \\
17:00:52.3 & 78:41:21 & 17157 & 37 & Abs & 17325 & 39 \\
17:00:59.2 & 78:44:59 & 15966 & 42 & Abs & 16137 & 53 \\
17:01:00.2 & 79:05:09 & 18377 & 42 & Emi & 18234 & 48 \\
17:01:34.6 & 78:50:46 & 19531 & 30 & Emi & 19518 & 15 \\
17:01:49.5 & 79:02:16 & 19307 & 90 & Emi & 19303 & 60 \\
17:02:07.5 & 78:17:29 & 17760 & 32 & Emi & 18001 & 47 \\
17:02:12.3 & 78:51:33 & 19325 & 16 & Emi & 19421 & 59 \\
17:02:17.8 & 78:45:52 & 20086 & 99 & Abs & 19907 & 66 \\
17:02:18.6 & 78:46:03 & 19581 & 38 & Emi & 19643 & 44 \\
17:02:46.8 & 78:51:46 & 16638 & 30 & Abs & 16574 & 53 \\
17:03:02.9 & 78:35:56 & 16417 & 39 & Abs & 16303 & 100 \\
17:03:10.9 & 78:14:40 & 18028 & 34 & Emi & 17973 & 48 \\
17:03:25.4 & 78:20:38 & 18922 & 43 & Emi & 18886 & 48 \\
17:03:29.5 & 78:37:55 & 17565 & 54 & Abs & 17796 & 44 \\
17:03:34.5 & 78:50:55 & 18989 & 72 & Emi & 18943 & 49 \\
17:03:56.5 & 78:44:44 & 17082 & 44 & Abs & 17115 & 36 \\
17:04:04.2 & 78:33:23 & 17316 & 53 & Abs & 17355 & 53 \\
17:04:14.5 & 79:23:13 & 26291 & 32 & Emi & \nodata & \nodata \\
17:04:15.4 & 78:06:07 & 18107 & 90 & Abs & \nodata & \nodata \\
17:04:41.0 & 78:13:49 & 17956 & 58 & Abs & 18008 & 56 \\
17:04:48.2 & 78:38:29 & 19013 & 38 & Abs & 19094 & 49 \\
17:04:54.0 & 78:29:54 & 52284 & 60 & Abs & 52295 & 114 \\
17:05:11.9 & 78:48:14 & 19921 & 32 & Emi & 19942 & 48 \\
17:05:21.8 & 79:08:05 & 18825 & 42 & Emi & 18807 & 48 \\
17:05:24.2 & 78:22:00 & 17189 & 55 & Abs & 17322 & 36 \\
17:05:33.6 & 78:51:16 & 19511 & 46 & Emi & 19479 & 54 \\
17:06:00.8 & 79:06:43 & 17908 & 68 & Abs & 17927 & 47 \\
17:06:28.4 & 78:49:15 & 18994 & 22 & Emi & 19093 & 45 \\
17:07:00.6 & 78:41:23 & 16956 & 62 & Abs & 17161 & 79 \\
17:07:02.4 & 79:10:19 & 11408 & 31 & Emi & 11436 & 49 \\
17:07:14.1 & 78:53:18 & 16563 & 36 & Abs & 16524 & 42 \\
17:07:15.9 & 78:30:08 & 18379 & 30 & Emi & 18582 & 48 \\
17:07:21.4 & 78:53:38 & 16520 & 56 & Abs & 16564 & 35 \\
17:08:33.5 & 78:29:54 & 18663 & 39 & Abs & 18684 & 46 \\
17:08:49.0 & 79:14:31 & 16668 & 44 & Abs & \nodata & \nodata \\
17:08:58.2 & 78:50:15 & 16249 & 99 & Abs & 16064 & 51 \\
17:09:10.0 & 78:52:32 & 18624 & 60 & Emi & 18812 & 46 \\
17:09:44.3 & 78:06:06 & 18039 & 32 & Emi & \nodata & \nodata \\
17:10:14.4 & 78:39:25 & 16922 & 47 & Abs & 17043 & 53 \\
17:10:15.9 & 78:02:07 & 17801 & 38 & Emi & \nodata & \nodata \\
17:10:31.5 & 78:29:10 & 15783 & 65 & Abs & 15792 & 66 \\
17:10:48.8 & 78:32:18 & 18682 & 36 & Abs & 18767 & 31 \\
17:13:57.2 & 78:59:33 & 19696 & 30 & Abs & 19614 & 62 \\
\enddata

\end{deluxetable}

\begin{deluxetable}{l l l r l l r l l c c}
\tablecolumns{11}
\tabletypesize{\small}
\tablecaption{Cluster Radio Galaxies\label{tbl-clusRG}}
\tablewidth{447pt}
\tablenum{2}
\tablehead{
\colhead{RA} & \colhead{Dec} & \colhead{$cz$} & 
\colhead{$\delta cz$} & \colhead{Source} & \colhead{$m_{R_c}$} & 
\colhead{$S_{1.4}$} & \colhead{$\delta S_{1.4}$} & \colhead{Sep} &
\colhead{ID} & \colhead{Class} \\
\multicolumn{2}{c}{(J2000)} & 
\multicolumn{2}{c}{[km s$^-1$]} & \colhead{} & \colhead{} &
\multicolumn{2}{c}{[mJy]} & \colhead{[\arcsec]} & \colhead{} & 
\colhead{}
}
\startdata
16:52:52.8 & 78:46:29 & 16805   & 27      & Ke      & 16.39 & 0.572 & 0.107 & 2.2 & \nodata & SF      \\  
16:53:05.8 & 78:30:13 & 16663   & 69      & Me      & 16.51 & 0.534 & 0.151 & 2.7 & \nodata & SF      \\
16:54:58.9 & 79:08:45 & 17433   & 61      & Mf      & 15.38 & 0.338 & 0.127 & 1.3 & \nodata & AGNo    \\
16:55:36.9 & 79:05:10 & 17808   & 30      & Kf      & 14.76 & 2.897 & 0.100 & 0.2 & \nodata & AGNo    \\
16:56:18.3 & 79:09:29 & 17616   & 36      & Ke      & 15.86 & 0.641 & 0.155 & 0.8 & \nodata & SF      \\
16:58:18.5 & 78:29:34 & 17882   & 47      & B       & 14.99 & 3.397 & 0.340 & 0.8 & AD & \nodata \\
16:58:46.8 & 78:25:21 & 15632   & 62      & Mf      & 15.23 & 1.000 & 0.218 & 5.0 & \nodata & AGNo    \\
17:00:25.8 & 78:49:26 & 15376   & 52      & B       & 16.58 & 0.225 & 0.071 & 1.5 & \nodata & \nodata \\
17:00:52.3 & 78:41:21 & 17157   & 37      & Mf      & 15.40 & 7.8\phn\phn & 0.3   & \tablenotemark{a} & I & AGNo \\
17:00:59.2 & 78:44:59 & 15966   & 42      & Kf Mf   & 15.29 & 0.247 & 0.060 & 0.4 & \nodata & AGNo    \\
17:01:00.2 & 79:05:09 & 18232   & 48      & Ke      & 16.20 & 0.865 & 0.124 & 0.5 & \nodata & SF      \\
17:01:23.4 & 78:41:16 & 16217   & 67      & B       & 15.55 & 0.425 & 0.126 & 2.2 & \nodata & \nodata \\
17:01:34.6 & 78:50:46 & 19531   & 30      & Me      & 16.64 & 0.549 & 0.083 & 0.4 & \nodata & SF      \\
17:01:49.5 & 79:02:16 & 19342   & 60      & Ke      & 14.65 & 2.962 & 0.100 & 1.2 & \nodata & AGNl    \\
17:02:12.3 & 78:51:33 & 19325   & 16      & Ke Me   & 15.76 & 1.438 & 0.100 & 0.7 & \nodata & SF      \\
17:02:15.0 & 78:35:47 & 18757   & 39      & B       & 15.36 & 0.642 & 0.185 & 1.7 & \nodata & \nodata \\
17:02:18.6 & 78:46:03 & 19612   & 23      & Ke      & 15.52 & 1.687 & 0.104 & 1.2 & K & SF      \\
17:02:46.8 & 78:51:46 & 16638   & 30      & Kf      & 15.12 & 1.840 & 0.100 & 0.7 & AC & AGNo    \\
17:03:02.9 & 78:35:56 & 16417   & 39      & Mf      & 14.90 & 50.3\phn\phn & 0.7   & \tablenotemark{a} & B & AGNo \\
17:03:22.4 & 78:46:55 & 19723   & 45      & B       & 16.33 & 0.454 & 0.087 & 1.2 & \nodata & \nodata \\
17:03:25.4 & 78:20:38 & 18922   & 43      & Me      & 16.43 & 0.714 & 0.155 & 0.6 & \nodata & SF      \\
17:03:29.5 & 78:37:55 & 17565   & 54      & Mf      & 14.73 & 120.6\phn\phn & 0.3   & \tablenotemark{a} & A & AGNo \\
17:03:30.1 & 78:39:55 & 17558   & 36      & B       & 15.01 & 40.7\phn\phn & 0.5   & \tablenotemark{a} & C & \tablenotemark{b} \\
17:03:34.1 & 78:37:48 & 16929   & 47      & B       & 14.70 & 1.261 & 0.148 & 2.3 & \nodata & \tablenotemark{b} \\
17:03:34.5 & 78:50:55 & 18989   & 72      & Me      & 16.02 & 0.697 & 0.115 & 1.4 & \nodata & SF      \\
17:03:56.5 & 78:44:44 & 17082   & 44      & Mf      & 15.51 & 5.816 & 1.000 & 3.1 & G & AGNo    \\
17:04:15.4 & 78:06:07 & 18107   & 90      & Kf      & 15.74 & 3.583 & 0.200 & 0.4 & \nodata & AGNo    \\
17:04:48.2 & 78:38:29 & 19103   & 38      & Mf      & 14.64 & 11.4\phn\phn & 0.1   & 1.9 & D & AGNo    \\
17:05:11.9 & 78:48:14 & 19921   & 32      & Me      & 16.20 & 0.576 & 0.112 & 0.3 & \nodata & SF      \\
17:05:33.6 & 78:51:16 & 19511   & 46      & Me      & 15.75 & 4.178 & 0.100 & 0.3 & E & SF      \\
17:06:43.4 & 78:35:57 & 14910   & 77      & B       & 16.17 & 1.025 & 0.076 & 0.9 & \nodata & \nodata \\
17:06:56.4 & 78:41:09 & 16867   & 38      & B       & 15.30 & 0.883\tablenotemark{c} & 0.104 & 1.7 & F3 & \nodata \\
17:07:15.9 & 78:30:08 & 18391   & 36      & Ke      & 16.41 & 1.329 & 0.086 & 1.2 & \nodata & SF      \\
17:08:33.5 & 78:29:54 & 18663   & 39      & Kf Mf   & 14.78 & 2.895 & 0.100 & 1.0 & AA & AGNo    \\
17:08:47.7 & 78:25:56 & 16860   &  7      & B       & 16.60 & 0.283 & 0.107 & 2.3 & \nodata & \nodata \\
17:08:58.1 & 78:50:15 & 16249   & 99      & Kf      & 15.21 & 0.600 & 0.083 & 1.2 & \nodata & AGNo    \\
17:09:10.0 & 78:52:32 & 18624   & 60      & Me      & 15.31 & 1.211 & 0.134 & 0.9 & \nodata & SF      \\
17:09:19.1 & 78:30:16 & 16700   & 44      & B       & 16.78 & 0.647 & 0.165 & 4.1 & \nodata & \nodata \\
17:09:39.9 & 78:33:07 & 16120   & 33      & B       & 15.44 & 0.500 & 0.132 & 2.9 & \nodata & \nodata \\
17:10:15.9 & 78:02:07 & 17801   & 38      & Me      & 16.58 & 0.405 & 0.105 & 1.7 & \nodata & SF      \\
           &          &         &         &         &       &       &       &     &         &         \\
           &          &         &         &         &       &       &       &     &         &         \\
           &          &         &         &         &       &       &       &     &         &         \\
           &          &         &         &         &       &       &       &     &         &         \\
           &          &         &         &         &       &       &       &     &         &         \\
           &          &         &         &         &       &       &       &     &         &         \\
\enddata

\tablenotetext{a}{Extended radio galaxy. See Figures \ref{fig-koalatad}--\ref{fig-flag}.}
\tablenotetext{b}{As a head-tail radio galaxy (see Figure \ref{fig-flagpole}), this galaxy may safely be considered an AGN.}
\tablenotetext{c}{See Figure \ref{fig-f}. The reported flux is only that associated directly with the galaxy; significantly more flux is associated with the ultrasteep spectrum source.}

\tablecomments{Sources for velocities are coded as follows: K -- Kitt Peak data; M -- MX data; B -- BLC02 velocity; f -- cross correlation velocity; e -- emission line velocity. When both KPNO and MX spectra were available, the velocities are combined through weighting by their errors. The $R_c$ magnitudes were calculated for the Gunn-Oke aperture, and have an associated error of $\lesssim0.05$ mags. The classes are determined from the galaxy spectra, and are defined as follows: SF -- star forming galaxy; AGNl -- AGN with a LINER spectrum; AGNo -- AGN with absorption line spectrum, sometimes with weak [NII] emission.}

\end{deluxetable}

\begin{deluxetable}{l l r r l l r l l c c c}
\tablecolumns{12}
\tabletypesize{\small}
\tablecaption{Additional Radio Galaxies\label{tbl-moreRG}}
\tablewidth{480pt}
\tablenum{3}
\tablehead{
\colhead{RA} & \colhead{Dec} & \colhead{$cz$} & 
\colhead{$\delta cz$} & \colhead{Source} & \colhead{$m_{R_c}$} & 
\colhead{$S_{1.4}$} & \colhead{$\delta S_{1.4}$} & \colhead{Sep} &
\colhead{ID} & \colhead{Class} & \colhead{Notes} \\
\multicolumn{2}{c}{(J2000)} & 
\multicolumn{2}{c}{[km s$^-1$]} & \colhead{} & \colhead{} &
\multicolumn{2}{c}{[mJy]} & \colhead{[\arcsec]} & \colhead{} & \colhead{} & \colhead{}
}
\startdata
16:53:19.4 & 78:30:40 & \nodata & \nodata & \nodata & 15.08 & 0.261 & 0.091 & 1.0 & \nodata & \nodata & \tablenotemark{a} \\
16:54:16.6 & 79:11:20 & 53488   & 53      & Mf      & 16.72 & 0.535 & 0.162 & 2.4 & \nodata & AGNo    & \tablenotemark{b} \\
16:54:36.0 & 79:14:56 & 39573   & 60      & Ke      & 16.72 & 0.520 & 0.145 & 1.4 & \nodata & SF      & \tablenotemark{b} \\
16:56:06.9 & 79:23:28 & 26352   & 60      & Ke      & 15.75 & 2.105 & 0.2   & 1.1 & \nodata & SF      & \tablenotemark{b} \\
16:57:58.5 & 78:36:15 & 4260    & 22      & Ke Me   & 15.22 & 1.922 & 0.1   & 0.3 & AF      & SF      & \tablenotemark{c} \\
17:02:57.9 & 78:03:18 & \nodata & \nodata & \nodata & 16.80 & 0.656 & 0.110 & 0.8 & \nodata & \nodata & \tablenotemark{a} \\
17:03:06.8 & 78:11:43 & \nodata & \nodata & \nodata & 16.79 & 0.429 & 0.141 & 3.0 & \nodata & \nodata & \tablenotemark{a} \\
17:04:14.5 & 79:23:13 & 26291   & 32      & Me      & 16.67 & 0.853 & 0.141 & 0.8 & \nodata & SF      & \tablenotemark{b} \\
17:04:54.0 & 78:29:54 & 52284   & 60      & Kf      & 16.64 & 1.898 & 0.154 & 1.5 & \nodata & AGNo    & \tablenotemark{b,d} \\
17:04:55.4 & 78:51:08 & 19980   & 45      & B       & 17.15 & 0.438 & 0.094 & 0.9 & \nodata & \nodata & \tablenotemark{e} \\
17:04:56.9 & 78:43:06 & 18682   & 45      & B       & 17.05 & 0.419 & 0.096 & 1.2 & \nodata & \nodata & \tablenotemark{e} \\
17:06:28.4 & 78:49:15 & 18994   & 22      & Me      & 16.96 & 1.135 & 0.098 & 1.4 & \nodata & SF      & \tablenotemark{e} \\
17:07:02.4 & 79:10:19 & 11408   & 31      & Ke Me   & 16.06 & 0.678 & 0.132 & 0.8 & \nodata & SF      & \tablenotemark{c} \\
17:07:40.1 & 78:56:57 & 53835   & 61      & B       & 17.57 & 0.276 & 0.078 & 1.6 & \nodata & \nodata & \tablenotemark{b} \\
17:09:01.1 & 78:46:18 & 11844   & 26      & B       & 15.24 & 2.180 & 0.1   & 0.9 & AE & \nodata & \tablenotemark{c} \\
17:11:20.7 & 78:49:47 & 39493   & 84      & B       & 16.64 & 1.231 & 0.1   & 1.0 & \nodata & \nodata & \tablenotemark{b} \\
17:14:18.9 & 78:05:20 & \nodata & \nodata & \nodata & 15.10 & 2.390 & 0.270 & 0.8 & \nodata & \nodata & \tablenotemark{a} \\
17:18:38.8 & 78:13:38 & \nodata & \nodata & \nodata & 15.90 & 1.299 & 0.100 & 1.0 & \nodata & \nodata & \tablenotemark{a} \\
\enddata

\tablenotetext{a}{Potential cluster radio galaxy.}
\tablenotetext{b}{Background radio galaxy.}
\tablenotetext{c}{Foreground radio galaxy.}
\tablenotetext{d}{See Figure \ref{fig-backfr1}.}
\tablenotetext{e}{Cluster radio galaxy, magnitude fainter than formal limit.}

\tablecomments{The $R_c$ magnitudes were calculated for the Gunn-Oke aperture, and have an associated error of $\lesssim0.05$ mags. The classes are determined from the galaxy spectra, and are defined as follows: SF -- star forming galaxy; AGNo -- AGN with optical spectrum dominated by an old stellar population, sometimes with weak emission of [NII]; }

\end{deluxetable}

\clearpage

\begin{deluxetable}{l r r r r r r r}
\tablecolumns{8}
\tablecaption{KMM Results\label{tbl-kmm}}
\tablewidth{0pt}
\tablenum{4}
\tablehead{
\colhead{} & \colhead{RA} & \colhead{Dec} &
\colhead{$N$} & \colhead{$N_{RG}$} &
\colhead{$N_{RG}$} & \colhead{$\bar v$} & \colhead{$\sigma$} \\
\colhead{Subcluster} & \colhead{(J2000)} & \colhead{(J2000)} &
\colhead{(total)} & \colhead{(cut)} &
\colhead{(full)} & \colhead{[\kms]} & \colhead{[\kms]}
}
\startdata
1 & 17 02 19 & 78 38 54 & 78  & 9  & 9  & 15895 & 581 \\
2 & 17 04 01 & 78 39 30 & 186 & 24 & 26 & 17790 & 704 \\
3 & 17 04 19 & 78 45 25 & 30  & 7  & 8  & 19708 & 278 \\
\enddata

\tablecomments{In the text, subcluster 2 is referred to as the ``Primary,'' subcluster 1 is referred to as the ``Subcluster,'' and subcluster 3 is referred to as the ``Group.''}
\end{deluxetable}

\clearpage

\begin{figure}
\figurenum{1}
\caption{The 1.4GHz image of the central 20\arcmin{} of Abell 2256, created using the multi-resolution clean algorithm to account for the extended emission. The resolution is 6.3\arcsec, and the rms noise is under 40 $\mu$Jy beam$^{-1}$. Alphabetical designations of prominent sources, assigned in prior studies, have been included. The radio relic is composed of the diffuse emission labeled ``G'' and ``H,'' while the ultrasteep spectrum source is associated with the ``F'' components. The cluster halo, not visible in this image, is approximately centered on source ``D.''\label{fig-bigmap}}
\end{figure}

\begin{figure}
\figurenum{2}
\caption{The tailed radio galaxies J170330+783755 and J170303+783556, sources ``A'' and ``B'' in BF76, respectively. Note that there are three galaxies in the system including J170330+783755, with the middle one also being a radio source. This radio source is likely an AGN, based on its elliptical host and its detection as a point source by {\it Chandra} (SMMV02). Contours are plotted at -3, 3, 5, 8, 13, 21, and 34 times the base level, which is 40 $\mu$Jy beam$^{-1}$.\label{fig-koalatad}}
\end{figure}

\begin{figure}
\figurenum{3}
\caption{The tailed radio galaxy J170330+783955, source ``C'' in BF76. Contours are plotted at -3, 3, 5, 8, 13, 21, and 34 times the base level, which is 40 $\mu$Jy beam$^{-1}$.\label{fig-flagpole}}
\end{figure}

\begin{figure}
\figurenum{4}
\caption{Completeness of velocity sampling, read as the fraction of galaxies inside a given radius for which velocity measurements are available.\label{fig-vcomp}}
\end{figure}

\begin{figure}
\figurenum{5}
\caption{The tailed radio galaxy J170052+784121, source ``I'' in BF76. Contours are plotted at -3, 3, 5, 8, 13, 21, and 34 times the base level, which is 40 $\mu$Jy beam$^{-1}$.\label{fig-flag}}
\end{figure}

\begin{figure}
\figurenum{6}
\caption{The ultrasteep spectrum source, designated as ``F'' in BF76. Contours are plotted at -3, 3, 5, 8, 13, 21, and 34 times the base level, which is 38 $\mu$Jy beam$^{-1}$. The filamentary structure is similar to that seen in the Abell 85 relic source, depicted in Figure 3 of \citet{slee2001}.\label{fig-f}}
\end{figure}

\begin{figure}
\figurenum{7}
\caption{The fairly strong radio galaxy J170448+783829, source ``D'' in BF76. The cluster halo source is approximately centered on this radio galaxy. Contours are plotted at -3, 3, 5, 8, 13, 21, and 34 times the base level, which is 40 $\mu$Jy beam$^{-1}$. The galaxy to the West is the brightest cluster galaxy.\label{fig-center}}
\end{figure}

\begin{figure}
\figurenum{8}
\caption{This galaxy resides at the peak in the emission associated with the Eastern edge of the relic, source ``G'' in BF76. Its large flux and error in Table \ref{tbl-clusRG} result from the attempt to fit the broad diffuse emission with a Gaussian. Contours are plotted at -3, 3, 5, 8, 13, 21, and 34 times the base level, which is 53 $\mu$Jy beam$^{-1}$.\label{fig-relic1}}
\end{figure}

\begin{figure}
\figurenum{9}
\caption{The radio galaxy J170150+790216, whose spectrum is that of a LINER. The optical flux peaks at the location of the radio emission. The rms noise is 59.0 $\mu$Jy beam$^{-1}$, with contours at -3, 3, 5, 8, 13, 21, and 34.\label{fig-fig8}}
\end{figure}

\begin{figure}
\figurenum{10}
\caption{The radio galaxy J170454+782954, which has a recession velocity of 52284 \kms. Its radio luminosity and morphology are that of an FR I source. No velocity is available for the strong radio galaxy to the East. The rms noise is 36.8 $\mu$Jy beam$^{-1}$, with contours at -3, 3, 5, 8, 13, 21, and 34.\label{fig-backfr1}}
\end{figure}

\begin{figure}
\figurenum{11}
\caption{Locations of the radio galaxies in Abell 2256. Stars represent star-forming galaxies, circles are AGN with absorption line spectra, triangles are LINERs, diamonds are confirmed cluster members for which no spectral classification was possible, and crosses are potential cluster radio galaxies which currently lack velocity measurements.\label{fig-dist}}
\end{figure}

\begin{figure}
\figurenum{12}
\caption{The locations of galaxies belonging to the identified group, plotted on the NVSS radio image centered on the relic. Members of the group are plotted as crosses, with additional circles indicating the group members which are also radio sources (all of which are powered by star formation). The three triangles are radio galaxies associated with the primary cluster but at only slightly higher probability than their being associated with the group. There are only six additional identified group members whose positions lie outside of this plot; only one is a radio source, and it is the LINER depicted in Figure \ref{fig-fig8}.\label{fig-relicpos}}
\end{figure}

\begin{figure}
\figurenum{13}
\caption{Difference between galaxy velocity and mean velocity of each substructure, plotted as a function of radial separation from the mean RA and Dec of the indicated substructure.\label{fig-trump}}
\end{figure}

\end{document}